\documentclass[%
 reprint,
 amsmath,amssymb,
 aps,
]{revtex4-2}

\usepackage{graphicx}
\usepackage{dcolumn}
\usepackage{bm}
\usepackage{dsfont}
\usepackage{braket}

\begin{document}

\preprint{APS/123-QED}

\title{Relaxing the i.i.d. state preparation assumption with the block QKD protocol}

\author{Olgierd T. \.{Z}urek}
 \email{Current email adress: olgierd.zurek@sorbonne-universite.fr}
\affiliation{Department of Physics, University of Toronto, Toronto, Canada}%

\author{Hoi-Kwong Lo}
\affiliation{Department of Electrical and Computer Engineering, University of Toronto, Toronto, Canada}%
\affiliation{
Center for Quantum Information and Quantum Control, University of Toronto, Toronto, Canada
}%
\affiliation{%
 Department of Physics, University of Toronto, Toronto, Canada}%
 \affiliation{%
 Centre for Quantum Technologies (CQT), Singapore}%
 \affiliation{%
 Department of Physics, National University of Singapore, Singapore}%

    \begin{abstract}
    In this work, we propose a novel approach to handling the independent and identically distributed assumption for the state preparation in prepare-and measure QKD protocols. Contrary to previous approaches, we do not prove the security using sub-protocols, and instead opt to guarantee security in the presence of correlations by dividing the communication into finite blocks of consecutive pulses, and adding an additional assumption of no dependence between each block. In turn we obtain a protocol which uses only a relaxed i.i.d. assumption and is easier to implement than subprotocol-style approaches since it does not require separate statistics for each sub-protocol, whose number grows with the correlation length of the underlying dependency.
\end{abstract}

\maketitle

\section{Introduction}
Quantum Key Distribution, or QKD, aims to solve the age-old problem of enabling secure communication between distant parties  \cite{Bennett_2014,ekert91}. Contrary to modern public-key cryptographic methods that rely on one-way hardness assumptions, QKD claims to be unconditionally secure against an eavesdropping attacker thanks to the laws of Quantum Mechanics. 

From its inception in 1984 by Bennet and Brassard \cite{Bennett_2014}, there have been multiple attempts to prove the security of QKD, notably by Mayers \cite{Mayers}, Lo and Chau \cite{Lo_1999} and Shor and Preskill \cite{Shor_2000}. Subsequent methods aimed to expand the proofs by considering imperfections within devices used to create and measure quantum states exchanged between the two parties. With the advent of Measurement Device Independent (MDI) QKD \cite{MDI_QKD}, all imperfections or Trojan horse attacks on the detector side have been taken into account by the security proofs. This left only the source-side imperfections as an argument for undermining the security of QKD.

Multiple methods have been developed to deal with fully and partially characterized imperfections in the source. For example, in the Loss-Tolerant Protocol \cite{Loss_tolerant}, the authors consider a QKD system using known arbitrary two-dimensional quantum states, proving the security without the symmetry properties needed in earlier proofs.  A recent example is a paper by Currás-Lorenzo et al \cite{Curras-Lorenzo_2025} that provides a convenient method to determine the security of QKD with a partially characterized source by using semi-definite optimization to determine the key rate. Nevertheless, a persistent assumption in all of these works was the independence and identical distribution (i.i.d.) of the emitted states within each round of the QKD protocol. This assumption is in principle unrealistic, and in order to claim unconditional security, one should first be able to certify that correlated imperfections do not pose a significant threat.

The area of non-i.i.d. security proofs has been so far underdeveloped. Some experimental works have attempted to gauge and model the magnitude of intensity correlations in decoy state methods \cite{Yoshino_2018,model_correlations_vigo} and correlations in phase encoding \cite{Amita_correlations}. However, those works were so far unable to quantify their findings using security frameworks that could take such imperfections into account, which frequently leads to assuming an intuitive yet unproven rule of thumb that low magnitude of correlations corresponds to a low impact on security. In 2020, a paper by Pereira et al \cite{corr2020} claimed to finally account for bounded correlations by separating a key distribution protocol into multiple sub-protocols, each one consisting only of states uncorrelated with each other. This approach was ultimately dubbed unsuccessful, since the authors did not consider the security of the composition of the keys obtained by each sub-protocol. Subsequent work addressed this issue by solving the composability problem by focusing on computing the phase error rates of each sub-protocol \cite{corr2026}. Additionally, recently another approach \cite{corr2026_2}, which relies on the quantum coin idea to prove security under correlations, was presented; however, at the time of writing this paper it is only available as a preprint. This shows that the topic of correlations in QKD is becoming more active, yet still there is no universally accepted security framework that covers correlations in state preparation.

In this work, we decide to explore a different avenue in the search for a non-i.i.d. security proof. In order to get around the i.i.d. assumption, we consider a new approach in which all correlations between subsequent states are encapsulated within finite blocks, which themselves are not correlated with each other. In this way, we are permitted to reuse existing security proof techniques, while allowing most of the correlations to be considered by the proof. Since our approach does not rely on the introduction of sub-protocols, Alice and Bob are only required to obtain average measurement statistics across the protocol, and do not have to compute them separately for each sub-protocol. This has the added benefit that the same class of protocols can be used with different correlation models, with the only requirement being that the dependency is erased after a block of a given number of pulses. Such erasure could be in principle performed using experimental techniques, for example by introducing bursting \cite{reem} for the case of correlations caused by bandwidth limitations \cite{Amita_correlations}.
\section{Block Protocol}

Let us start by imagining all of the states that Alice sends to Bob throughout a discrete-variable quantum key distribution protocol, assuming that Alice's source introduces a non-i.i.d. imperfection. For simplicity, we consider only classical correlation between states, and also assume that the states are pure. In general, we can describe these states as
\begin{equation}
    \ket{\phi_{total}} = \ket{\psi_{i_1}}\otimes\ket{\psi_{i_2|i_1}}\otimes\ket{\psi_{i_3|i_2,i_1}}\otimes \ldots
\end{equation}
where $\ket{\psi_{i_1}}$ represents the state send by Alice in the first round, in which she decided on the setting $i_1 \in \{0_\mathcal{Z},1_\mathcal{Z},0_\mathcal{X},\ldots\}$ out of the settings used in the particular protocol she is running. Her subsequent state $\ket{\psi_{i_2|i_1}}$ depends not only on her second round setting choice $i_2$, but also on $i_1$, making this a non-i.i.d. state preparation method. In principle, dependencies like this can be unbounded, though their exact form depends on the preparation process in the source. 

In our model, we artificially remove any dependence every $m$ states, effectively erasing the memory of Alice's source. This will modify our equation from above as follows
\begin{equation}
\begin{gathered}
    \ket{\phi_{block}} = \ket{\psi_{i_1}}\otimes\ket{\psi_{i_2|i_1}}\otimes \ldots \otimes \ket{\psi_{i_m|i_{m-1}\ldots i_1}}\otimes \\\otimes\ket{\psi_{i_{m+1}}}\otimes\ket{\psi_{i_{m+2}|i_{m+1}}} \otimes \ldots
\end{gathered}
\end{equation}

Let us call the states $m$ labeled with $i_1\ldots i_m$ as a block. In our model we allowed the correlations to be exhibited within each block, however, there are no correlations between blocks. This allows us to treat all blocks as i.i.d. with respect to each other. This property in particular allows us to use i.i.d. methods to determine the security of the key created using this block model.

We formalize the steps of the protocol below:
\begin{enumerate}
    \item Alice chooses to send a key generation block or a parameter estimation block with some probability $p_{key}$. If she chooses the key generation block, she prepares a $m$-pulse block  of $\mathcal{Z}$ basis states, where the bit values are chosen randomly with equal probability. If she chooses the parameter estimation block, she prepares a block by selecting an m-pulse state $\ket{\psi_{\vec{i}}}$ with probability $P(\vec{i})$. She sends the prepared pulse to Bob.
    \item Bob selects the basis for his measurement of the block as either the $\mathcal{X}$ basis or the $\mathcal{Z}$ basis. He then uses the same basis for measuring all the pulses in the block.
    \item Alice announces her block type choice for each round and Bob announces his basis choice.
    \item The rounds in which Alice send a key generation block and Bob measured in the $\mathcal{Z}$ basis are called key generation rounds. The rounds where Alice chose the  parameter estimation block and Bob selected the $\mathcal{X}$ basis are dubbed parameter estimation rounds.
    \item Alice and Bob share part of the measurement results from the key generation rounds to compute the QBER
    \item Using parameter estimation rounds, Alice and Bob compute the probabilities of obtaining a measurement result $\gamma$ in the $\mathcal{X}$ basis, given that for that specific pulse a setting $i$ was selected by Alice. They use this data to upper bound the phase error rate.
    \item They perform error correction and privacy amplification on setting choices and measurement results from the key generation rounds.
\end{enumerate}

To compute the key rate, we will use a source replacement scheme \cite{Lo_1999,Shor_2000}. We are interested in the phase error rate of the virtual state of the entire block in the key generation rounds, given by
\begin{equation}
    \ket{\Psi_{\mathcal{Z}}^{\otimes m}} = \frac{1}{\sqrt{2^{m}}}\sum_{\vec{i}\in \{0,1 \}^{\otimes m}}\ket{\vec{i}}\ket{\psi_{\vec{i}}}
\end{equation}
We can think of the virtual key distribution protocol as attempting to distill some number of perfect EPR pairs from the $m$ noisy EPR pairs given by the above equation \cite{Shor_2000}. In order to calculate the number of perfect EPR pairs that can be distilled we need to find the average phase error rate of the block. Following the approach of Curras-Lorenzo et al \cite{Curras-Lorenzo_2025}, we opt to use a convex optimization program, since it allows for flexibility when dealing with different imperfection models. 

Since we treat the block as a set of noisy EPR pairs, the overall phase error rate will be the average of the phase error rates for each pulse in the block, weighted by the probability that a given pulse is measured by Bob. A phase error rate for pulse $k$ is given by
\begin{equation}
\begin{gathered}
e_{ph}^k = \frac{Y_{\mathcal{Z}\land ph}^k}{Y^k_{\mathcal{Z}}}\\    
    Y_{\mathcal{Z}\land ph}^k = \textrm{tr}(\mathds{1}_{A}\otimes\Lambda_{B}(\ket{\Psi_{\mathcal{Z}}^{\otimes m}}\bra{\Psi_{\mathcal{Z}}^{\otimes m}})\cdot \\\cdot(\hat{P}_{0_{\mathcal{X}}}^k\otimes\hat{\Gamma}_{1_{\mathcal{X}}}^k+\hat{P}_{1_{\mathcal{X}}}^k\otimes\hat{\Gamma}_{0_{\mathcal{X}}}^k))
\end{gathered}
\end{equation}
where $\hat{P}_x^k$ is the projection onto the subspace $\ket{x}\bra{x}$ of the $k$-th Alice's qubit, $\hat{\Gamma}^k_{\gamma}$ is the POVM matrix of Bob's $\gamma$ result on the $k$-th pulse, $\mathds{1}_{A}\otimes\Lambda_{B}$ is a coherent attack on the entire block and $Y^k_{\mathcal{Z}}$ is the $\mathcal{Z}$ basis gain for pulse $k$. It is worth noting that, while $\hat{P}_x^k$ and $\hat{\Gamma}^k_{\gamma}$ are concerned with a single pulse, they are POVM matrices on the Hilbert space of $m$ qubits and $m$ pulses, respectively. The average phase error rate is then given by
\begin{equation}
    e_{ph} = \sum_k e_{ph}^k\frac{Y_{\mathcal{Z}}^k}{\sum _kY_\mathcal{Z}^k} = \frac{1}{mY_\mathcal{Z}}\sum_k Y^{k}_{\mathcal{Z} \land ph }
\end{equation}
where we needed to include the term $\frac{Y_{\mathcal{Z}}^k}{\sum _kY_\mathcal{Z}^k}$ to take into account the probability that Bob successfully measures pulse $k$ in the key generation round.
Let us assume that Eve's attack can be described using the following Kraus representation
\begin{equation}
    \mathds{1}_{A}\otimes\Lambda_{B}(\rho_{AB}) = \sum_l \mathds{1}\otimes K_l \rho_{AB} \mathds{1} \otimes K_l^{\dag}
\end{equation}
If we explicitly add this to the equation for $Y_{\mathcal{Z}\land ph}^k$, we get
\begin{equation}
\begin{gathered}
    Y_{\mathcal{Z}\land ph}^k = \textrm{tr}(\ket{\Psi_{\mathcal{Z}}^{\otimes m}}\bra{\Psi_{\mathcal{Z}}^{\otimes m}}\sum_l\mathds{1}\otimes K_l^{\dag}(\hat{P}_{0_{\mathcal{X}}}^k\otimes\hat{\Gamma}_{1_{\mathcal{X}}}^k+ \\ +\hat{P}_{1_{\mathcal{X}}}^k\otimes\hat{\Gamma}_{0_{\mathcal{X}}}^k)\mathds{1} \otimes K_l) =\\
    =\bra{\Psi_{\mathcal{Z}}^{\otimes m}}\hat{P}_{0_{\mathcal{X}}}^k\otimes \sum_lK_l^{\dag}\hat{\Gamma}_{1_{\mathcal{X}}}^kK_l+\\+\hat{P}_{1_{\mathcal{X}}}^k\otimes\sum_lK_l^{\dag}\hat{\Gamma}_{0_{\mathcal{X}}}^kK_l\ket{\Psi_{\mathcal{Z}}^{\otimes m}} =: \\
    =:\bra{\Psi_{\mathcal{Z}}^{\otimes m}}\hat{P}_{0_{\mathcal{X}}}^k\otimes \hat{Q}_{1_{\mathcal{X}}}^k+\hat{P}_{1_{\mathcal{X}}}^k\otimes \hat{Q}_{0_{\mathcal{X}}}^k\ket{\Psi_{\mathcal{Z}}^{\otimes m}} 
\end{gathered}
\end{equation}
where we defined a new POVM $\hat{Q}_{\gamma}^k = \sum_lK_l^{\dag}\hat{\Gamma}_{\gamma}^kK_l$. 

Now, if we knew the form of POVM matrices $\hat{Q}_{\gamma}^k$, we could derive the phase error rate of the block. In order to determine that form we will use a convex program with constraints coming from the measurement statistics obtained by Alice and Bob. The exact form of the constraints will depend on the model of the protocol and the amount of knowledge that we assume Alice and Bob can obtain.

In principle, Alice and Bob can obtain additional information about Eve's attack by randomly dedicating some blocks for parameter estimation. We assume that, for each block, there is a non-zero probability that Alice and Bob individually choose to perform a parameter estimation round instead of a key round. We assume that if a block is designated as a parameter estimation round, Alice chooses to send a random block state $\ket{\psi_{\vec{i}}}$ with probability $P(\vec{i})$, where $\vec{i}\in \{0_{\mathcal{Z}},1_{\mathcal{Z}},0_{\mathcal{X}}\ldots\}^{\otimes m}$, and Bob chooses the $\mathcal{X}$ basis measurements for each pulse in the block. For blocks labeled as key generation rounds, Alice randomly sends a state $\ket{\psi_{\vec{i}}}$ where $\vec{i}\in \{0_{\mathcal{Z}},1_{\mathcal{Z}}\}^{\otimes m}$ with equal probability over the set, and Bob chooses the $\mathcal{Z}$ basis for each pulse in the block.

The first constraint of $\hat{Q}_{\gamma}^k$ is a normalization condition, namely that, for each $k$ we have
\begin{equation}
    \sum_{\gamma}\hat{Q}_{\gamma}^k =\sum_lK_l^{\dag}\sum_{\gamma}\hat{\Gamma}_{\gamma}^kK_l = \sum_lK_l^{\dag}K_l = \mathds{1}
\end{equation}
Secondly, since $\Lambda$ is a completely positive map, we must have that for any $\rho$
\begin{equation}
\begin{gathered}
    \textrm{tr}(\rho \hat{Q}_{\gamma}^k) = \textrm{tr}(\rho \sum_lK_l^{\dag}\hat{\Gamma}_{\gamma}^kK_l) =\\= \textrm{tr}(( \sum_lK_l\rho K_l^{\dag})\hat{\Gamma}_{\gamma}^k) = \textrm{tr}(\Lambda(\rho) \hat{\Gamma}_{\gamma}^k) \geq 0
\end{gathered}
\end{equation}
which is always greater than or equal to 0 since $\hat{\Gamma}_{\gamma}^k$ as a POVM matrix is positive semi-definite. The above holds for all $\rho$, hence we must have that $\hat{Q}_{\gamma}^k$ is positive semi-definite.

Lastly, we have the constraints that come from Bob's $\mathcal{X}$ basis statistics. In this protocol, we assume that Alice and Bob only check the statistics averaged over the pulses in the block, and over the possible state choices by Alice. This is equivalent to assuming the knowledge of $P(\gamma|i)$, where $\gamma$ labels Bob's measurement result and $i$ indicates the setting choice by Alice.

The probability that Bob obtains the result $\gamma$ for the $k$-th pulse, given that we know that Alice chose the setting $i$ for that pulse, is given by
\begin{equation}
    P_k(\gamma|i) = \sum_{\vec{i}:i_k = i}P(\vec{i})\bra{\psi_{\vec{i}}}\hat{Q}_{\gamma}^k\ket{\psi_{\vec{i}}}/p_k(i)
\end{equation}
where $P(\vec{i})$ is the probability that the setting $\vec{i}$ is chosen by Alice, while $p_k(i)$ is the probability that the setting $i$ is chosen for the $k$-th pulse. Now we need to take the average over all $m$ pulses. This corresponds to the constraint
\begin{equation}
    Y^{\gamma}_i = \frac{1}{m}\sum_k \sum_{\vec{i}:i_k = i}P(\vec{i})\bra{\psi_{\vec{i}}}\hat{Q}_{\gamma}^k\ket{\psi_{\vec{i}}}/p_k(i)
\end{equation}

We have direct access to the operators needed to implement the average statistics constraints, therefore we can formulate the above as a convex optimization over the operators $\hat{Q}_{\gamma}^k \in \mathds{C}^{d\times d}$, where $d$ is the dimension of the Hilbert space of a single pulse. The explicit form of that program is then given by
\begin{equation}
\begin{aligned}
\underset{\hat{Q}_\gamma^k,\;k=1,\ldots m}{\text{maximize }}   &\frac{1}{m}\sum_{k = 1}^{m}\sum_{x \in \{0_{\mathcal{X}},1_{\mathcal{X}}\}}\bra{\Phi}\hat{O}_{\gamma}^k \ket{\Phi}/Y_{\mathcal{Z}} \\
\text{subject to } &  \frac{1}{m}\sum_k \sum_{\vec{i}:i_k = i}P(\vec{i})\bra{\psi_{\vec{i}}}\hat{Q}_{\gamma}^k\ket{\psi_{\vec{i}}}/p_k(i) = Y^{\gamma}_i\\
&\sum_\gamma \hat{Q}_{\gamma}^k = \mathds{1} \\
&\hat{Q}_{\gamma}^k \succeq 0 \textrm{  for all }\gamma, k, \vec{i}
\end{aligned}
\end{equation}
where the objective is the average of the individual phase error rates $e_{ph}^k$ weighted by the gain $Y^k_{\mathcal{Z}}$ of each pulse.

\section{Examples}
\begin{figure}[t]
    \centering
    \includegraphics[width=0.90\linewidth]{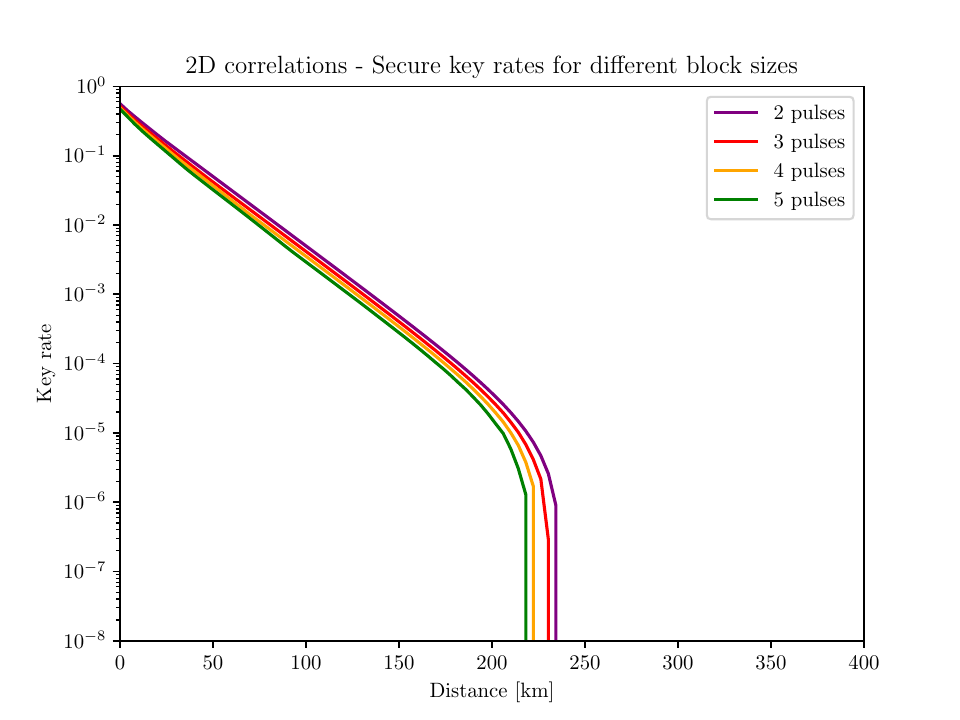}
\caption{Secure key rates for the case of 2D bit-and-basis next-neighbor correlations. Note that increasing the block size lowers the key rate, since we allow for more correlations to persist in the communication, therefore more leakage is available to Eve.}
    \label{figure_11}
\end{figure}

\paragraph*{2D Nearest-neighbor correlations}We start by presenting an example of a simple correlated qubit flaw. Here, we assume that the states exchanged by Alice and Bob are qubit sates, but their polarization can contain imperfections depending on the previous state choices. In the following model we assume only next-neighbor correlations, which are described by
\begin{equation}
\begin{gathered}
 \ket{\psi_{i_n|i_{n-1}}} = \begin{bmatrix}
\cos(\phi'_{i_n|i_{n-1}}/2) \\
\sin(\phi'_{i_n|i_{n-1}}/2) 
\end{bmatrix} \\
    \phi_{i_n|i_{n-1}}' = (1+\frac{\delta}{\pi}(-1)^{b_{n-1}}(1/2+x_{n-1}))\phi_{i_n},
\end{gathered}
\end{equation}
where $x_{n-1},b_{n-1} \in \{0,1\}$ are binary labels corresponding to the bit and basis choices of the state from the previous round, labeled with $n-1$, and $\phi_{i_n}$ is the ideal polarization to encode the $\ket{\psi_{i_n}}$ state in the absence of imperfections. Using this model, we can now construct the states of the entire block by writing for all $\vec{i}$
\begin{equation}
    \ket{\psi_{\vec{i}}} = \ket{\psi_{i_1}}\ket{\psi_{i_2|i_{1}}}\ldots\ket{\psi_{i_m|i_{m-1}}}
\end{equation}
where $\ket{\psi_{i_1}}$ is the perfect encoding where $\phi_{i_1}' = \phi_{i_1}$. While this model is not inspired by a physical system, it is valuable to examine since it offers a straightforward way to encode both bit and basis dependence.  The average key rates for this model, obtained for $\delta = 0.3$ and the dark count rate $p_d = 10^{-6}$ are presented in Figure \ref{figure_11}. We assumed here that the probability of choosing the $\mathcal{Z}$ basis by Bob and the key generation round for Alice is close to one, and that for the parameter estimation rounds Alice chooses every state combination with equal probability.  

We observe that increasing the block size causes a drop in the key rate. This is expected since for larger blocks we allow Eve to do a more sophisticated attack that encapsulates more pulses, while not adding additional constraints to the optimization problem, since Alice and Bob only check the average statistics.

\begin{figure}[t]
    \centering
    \includegraphics[width=0.9\linewidth]{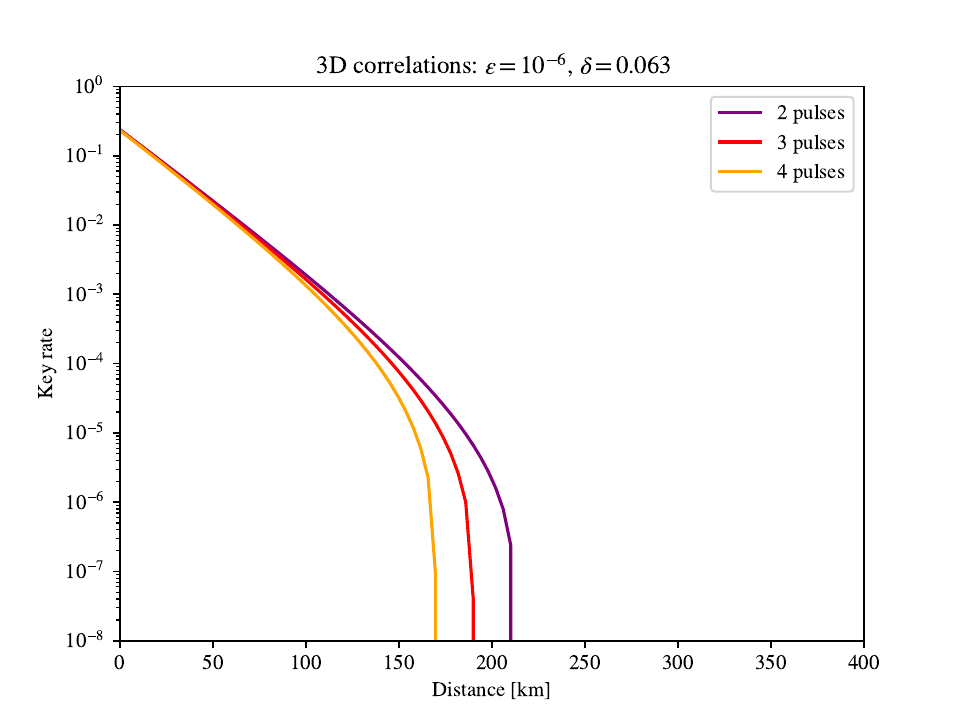}
    \caption{Average key rate for the correlation from equation \ref{model_2}, given the particular form described by equation \ref{particular}. Similarly to the previous case we observe that increasing the block size impacts the key rate. Nevertheless, a leakage parametrized by $\epsilon = 10^{-6}$ does not cause a significant drop in the key rate between the different block sizes. }
    \label{figure_2}
\end{figure}

\paragraph*{Side channels and correlations} As a second example we show a model inspired by a scenario from \cite{corr2020}. In that paper, in equations (14) and (15), the authors consider a partially-characterized nearest neighbor correlation model of the form
\begin{equation}
     \ket{\psi_{i_n|i_{n-1}}} = (1-\epsilon)\ket{\varphi_{i_n}} + \sqrt{1-(1-\epsilon)^2}\ket{\varphi^{\perp}_{i_n|i_{n-1}}}
\end{equation}
where $\ket{\phi_{i_n}}$ is one of the imperfect but characterized BB84 states
\begin{equation}
\begin{gathered}
 \ket{\varphi_{i_n}} =
\cos(\phi'_{i_n}/2)\ket{0}+\sin(\phi'_{i_n}/2) \ket{1}\\
    \phi_{i_n}' = (1+\frac{\delta}{\pi})\phi_{i_n},
\end{gathered}
\label{model_2}
\end{equation}
and $\ket{\varphi^{\perp}_{i_n|i_{n-1}}}$ is an uncharacterized state depending on the setting $i_{n-1}$, which is only known to be orthogonal to the states $\ket{\varphi_{i_n}}$ for all $i_n,i_{n-1}$.
In our approach we assume the states are fully characterized, hence we propose a following model for the state $\ket{\varphi^{\perp}_{i_n|i_{n-1}}}$:
\begin{equation}
    \ket{\varphi^{\perp}_{i_n|i_{n-1}}} = (-1)^{b_{n-1}}\ket{2}
    \label{particular}
\end{equation}
where the only dependence on the previous state is an overall phase change given by the basis choice of the previous pulse. Given this correlation and imperfection model, in Figure \ref{figure_2} we present the key rates for different block sizes, for $p_d = 10^{-7}$, $\delta = 0.063$ and $\epsilon = 10^{-6}$.

\section{Conclusions}
In this work we present a new and conceptually simple QKD framework that does not require the traditional i.i.d. state preparation assumption between signals in order to guarantee security. It is easy to implement experimentally, since one can create the uncorrelated blocks by turning off the laser for an extended period of time, and is also applicable to a wide range of correlation models, thanks to the use of convex programming to compute the key rate. It therefore offers a novel and viable approach to considering correlations in QKD and may be a stepping stone towards future protocols which do not need i.i.d. assumptions while remaining practical and yielding high key rates.

The block approach can also be considered with other knowledge assumptions for Alice and Bob. For example, one could assume that they calculate the $\mathcal{X}$ basis statistics for each pulse in the block separately, which may lead to better key rates since it puts more constraints on Eve. This and other related cases, such as considering periodic correlations within a block, were discussed in the master thesis of O\.{Z} \cite{master}. This research also showed numerically that under the more extensive knowledge assumptions and a restriction to 2D Hilbert space of prepared states, the key rate of the block follows a simple formula that can be extended to very large block sizes, which bypasses the requirement for high computing power.

Nevertheless, extending the result to full coherent attacks requires further research, since currently the attacker is limited only to coherent attacks on pulses within a single block. We can extend the attacker's capacity by increasing the block size, but this also modifies the protocol, hence the security against a true coherent attack remains an open problem.
\section{Acknowledgement}
We thank Kiyoshi Tamaki for helpful comments.

This project is supported by NSERC Discovery Grant. 
O\.Z thanks for the funding from the Erasmus+: Erasmus
Mundus programme of the European Union. HKL is supported by the funding from the National Research Foundation, Singapore, through the National Quantum Office, hosted in A*STAR, under its Centre for Quantum Technologies Funding Initiative (S24Q2d0009), and NUS Start-Up Grant A-0010223-00-00 and funding from CFI and ORF.
\vspace{30px}
\addcontentsline{toc}{section}{References}
\bibliographystyle{unsrt}
\bibliography{apssamp}

\section*{Appendix A - Derivation of the measurement statistics}
\addcontentsline{toc}{chapter}{Appendix D - Derivation of the measurement statistics}
\subsection*{Preparation of the states for a given model of imperfections and leakage}
In order to run our key rate simulation, we need to define a set of states $\{\ket{\psi_i}\}_i$ to be used within the model. For i.i.d. imperfections, each state in that set corresponds to a setting choice used in a given prepare-and-measure protocol, but the form of the prepared state is affected by the preparation flaw. In order to obtain statistics for a particular imperfection model, for each state, we define a two-dimensional encoding subspace within the Hilbert space to it belongs. We assume that Bob designs his measurements so that he only measures states that lie within that subspace, or equivalently, he starts his measurement by projecting the state onto that subspace, which is similar to the statement of the squash model \cite{Squash}. We would like to stress that this assumption is not required for the security of our approach and is only used to compute realistic statistics for the sake of benchmarking.

For the block model, the set of states $\{\ket{\psi_{\vec{i}}}\}_{\vec{i}}$ contains the states corresponding to every possible combination of setting choices by Alice for each pulse. Then, for every setting choice, we use the particular state preparation flaw and leakage model to compute the exact form of the state. At this point, we can also choose to implement the periodic boundary condition.
\subsection*{Calculation of the statistics}
In this section, we will describe the method to simulate the statistic used in the convex programs. We will do so by assuming a particular form of Bob's POVM matrices. 

For now let us assume that Bob's POVM is a two dimensional matrix, according to the ideal model of BB84 or six-state protocol. We know that, ideally, Bob's $\mathcal{X}$ measurements would correspond to 
\begin{equation}
    \hat{\Gamma}_{0_{\mathcal{X}}} = \ket{{0_{\mathcal{X}}}}\bra{{0_{\mathcal{X}}}}, \;\;\;\;\hat{\Gamma}_{1_{\mathcal{X}}} = \ket{{1_{\mathcal{X}}}}\bra{{1_{\mathcal{X}}}},\;\;\;\;\hat{\Gamma}_{f} = 0.
\end{equation}
Now, we need to take into account realistic imperfections of the QKD setup: the dark count rate $p_d$, detector efficiency $\eta_{det}$ and attenuation
coefficient $\alpha$ \cite{statistics}. The attenuation coefficient can be used to calculate the fiber transmission rate
\begin{equation}
    \eta_{channel} = 10^{-\alpha l}
\end{equation}
where $l$ is the length of the fiber. Combining the two coefficients $\eta :=\eta_{channel}\cdot \eta_{det}$ gives us the overall transmission efficiency. 

Given some $\eta$ and $p_d=0$, it is easy to see that the POVM will be modified to
\begin{equation}
\begin{gathered}
    \hat{\Gamma}_{0_{\mathcal{X}}} = \eta\ket{{0_{\mathcal{X}}}}\bra{{0_{\mathcal{X}}}}, \;\;\;\;\hat{\Gamma}_{1_{\mathcal{X}}} = \eta\ket{{1_{\mathcal{X}}}}\bra{{1_{\mathcal{X}}}},\\\hat{\Gamma}_{f} = (1-\eta) \mathds{1}.
\end{gathered}
\end{equation}
Now we incorporate the dark count rate into our model. The dark count rate is the probability that a detector produces a count without being triggered by external light. In the setup considered, Bob uses two detectors, each measuring one of the $\mathcal{X}$ basis states. If a tick is detected by both detectors then a result is chosen randomly between the two non-failure options. We need to therefore examine the following cases: 1) Light arrives at one of the detectors and no dark counts are detected; 2) Light arrives at one of the detectors and a dark count is detected by the same detector; 3) Light arrives at one of the detectors and a dark count is detected by the other detector; 4) Light arrives at one of the detectors and a dark count is detected by both detectors; 5) Light does not arrive and no dark counts are detected; 6) Light does not arrive and a dark count is detected by one of the detectors; 7) Light does not arrive and a dark count is detected by both detectors. For simplicity, we omit the higher order cases, where multiple dark counts are detected, and at the end we will round our result up to terms linear in $p_d$.

We can think of each of those cases as separate outcomes measured by Bob and construct a POVM whose results correspond to the above events, for each of the detectors. Starting with case 1), the event where light reaches the detector and is measures occurs with a probability $\eta$, while the event that no dark counts are detected is given by $(1-p_d)^2$, with the square coming from the fact that we have two detectors. Now, the POVM matrix for the case when  $0_{\mathcal{X}}$ is detected is simply a projection $\ket{{0_{\mathcal{X}}}}\bra{{0_{\mathcal{X}}}}$, while for detection of  $1_{\mathcal{X}}$ is given by $\ket{{1_{\mathcal{X}}}}\bra{{1_{\mathcal{X}}}}$. Renormalizing these matrices using the classical probability $(1-p_d)^2\eta$, we get that, for the events from case 1) and Bob's results $0_{\mathcal{X}}, 1_{\mathcal{X}},f$, the POVMs are given by
\begin{equation}
\begin{gathered}
    \hat{\Gamma}_{0_{\mathcal{X}}}^1 = (1-p_d)^2\eta\ket{{0_{\mathcal{X}}}}\bra{{0_{\mathcal{X}}}},\\\hat{\Gamma}_{1_{\mathcal{X}}}^1 = (1-p_d)^2\eta\ket{{1_{\mathcal{X}}}}\bra{{1_{\mathcal{X}}}},\;\;\;\;\hat{\Gamma}_{f}^1 = 0.
\end{gathered}
\end{equation}
Case 2) is similar, since the additional dark count does not influence the result. After modifying the classical probability to $(1-p_d)p_d\eta$ the POVMs for this case are equal to
\begin{equation}
\begin{gathered}
    \hat{\Gamma}_{0_{\mathcal{X}}}^2 = (1-p_d)p_d\eta\ket{{0_{\mathcal{X}}}}\bra{{0_{\mathcal{X}}}},\\\hat{\Gamma}_{1_{\mathcal{X}}}^2 = (1-p_d)p_d\eta\ket{{1_{\mathcal{X}}}}\bra{{1_{\mathcal{X}}}},\;\;\;\;\hat{\Gamma}_{f}^2 = 0,
\end{gathered}
\end{equation}
For case 3) we need to account for the double counts. The classical probability of the case occurring remains the same as for case 2), but now we need to take into account also that after the double count, Bob would randomly assign the result to two outcomes. Therefore, assuming that the light reached detector $0_{\mathcal{X}}$ and a dark count was observed by detector measuring $1_{\mathcal{X}}$, a POVM matrix corresponding to the $0_{\mathcal{X}}$, the POVM matrix should be of the form $\ket{{0_{\mathcal{X}}}}\bra{{0_{\mathcal{X}}}}$ for both successful events, renormalized by the classical probability $(1-p_d)p_d\eta$ of the dark count and measurement happening and also by $1/2$, coming from the random assignment of the result. By combining the contributions for both dark count/true count pairs, we get
\begin{equation}
\begin{gathered}
    \hat{\Gamma}_{0_{\mathcal{X}}}^3 = \frac{1}{2}(1-p_d)p_d\eta(\ket{{0_{\mathcal{X}}}}\bra{{0_{\mathcal{X}}}} + \ket{{1_{\mathcal{X}}}}\bra{{1_{\mathcal{X}}}}), \\\hat{\Gamma}_{1_{\mathcal{X}}}^3 = \frac{1}{2}(1-p_d)p_d\eta(\ket{{0_{\mathcal{X}}}}\bra{{0_{\mathcal{X}}}} + \ket{{1_{\mathcal{X}}}}\bra{{1_{\mathcal{X}}}}),\;\;\;\;\hat{\Gamma}_{f}^3 = 0.
\end{gathered}
\end{equation}
Case 4) is analogous to case 4), but we need to replace the probability of the event occurring with $\eta p_d^2$
\begin{equation}
\begin{gathered}
     \hat{\Gamma}_{0_{\mathcal{X}}}^4 = \frac{1}{2}p_d^2\eta\ket{{0_{\mathcal{X}}}}\bra{{0_{\mathcal{X}}}} + \frac{1}{2}p_d^2\eta\ket{{1_{\mathcal{X}}}}\bra{{1_{\mathcal{X}}}}, \\\hat{\Gamma}_{1_{\mathcal{X}}}^4 = \frac{1}{2}p_d^2\eta\ket{{0_{\mathcal{X}}}}\bra{{0_{\mathcal{X}}}} + \frac{1}{2}p_d^2\eta\ket{{1_{\mathcal{X}}}}\bra{{1_{\mathcal{X}}}},\;\;\;\;\hat{\Gamma}_{f}^4 = 0,
\end{gathered}
\end{equation}
Case 5) is the only failure case, which occurs when neither the pulse nor the dark counts are detected. The corresponding POVM is
\begin{equation}
    \hat{\Gamma}_{0_{\mathcal{X}}}^5 = 0, \;\;\;\;\hat{\Gamma}_{1_{\mathcal{X}}}^5 = 0,\;\;\;\;\hat{\Gamma}_{f}^5 = (1-p_d)^2(1-\eta)\mathds{1},
\end{equation}
In case 6), the light does not reach the detector, which has a $1-\eta$ probability of happening, and only one dark count is measured, which occurs with probability $p_d(1-p_d)$. Since the results are not contingent on the form of the incoming state, we simply renormalize the identity matrix by the classical probability, to obtain
\begin{equation}
\begin{gathered}
    \hat{\Gamma}_{0_{\mathcal{X}}}^6 = (1-\eta)p_d(1-p_d)\mathds{1}, \\\hat{\Gamma}_{1_{\mathcal{X}}}^6 =(1-\eta)p_d(1-p_d)\mathds{1},\;\;\;\;\hat{\Gamma}_{f}^6 =0.
\end{gathered}
\end{equation}
Finally for case 7), we have a similar situation to case 6), but now, the classical probability is given by $(1-\eta)p_d^2$. Also, since the results are randomized across the two options, the POVM matrices need to be multiplied by $1/2$. We have that
\begin{equation}
\begin{gathered}
    \hat{\Gamma}_{0_{\mathcal{X}}}^7 = \frac{1}{2}(1-\eta)p_d^2\mathds{1}, \\\hat{\Gamma}_{1_{\mathcal{X}}}^7 =\frac{1}{2}(1-\eta)p_d^2\mathds{1},\;\;\;\;\hat{\Gamma}_{f}^7 =0.
\end{gathered}
\end{equation}
We now sum all the matrices of this POVM corresponding to a given result, since Bob doesn't have the capacity to distinguish between them. We obtain that
\begin{equation}
    \begin{gathered}
     \hat{\Gamma}_{0_{\mathcal{X}}} = \eta(1-p_d/2)\ket{{0_{\mathcal{X}}}}\bra{{0_{\mathcal{X}}}} +\eta p_d/2\ket{{1_{\mathcal{X}}}}\bra{{1_{\mathcal{X}}}} +\\+ (1-\eta)p_d(1-p_d/2)\mathds{1}, \\\hat{\Gamma}_{1_{\mathcal{X}}} = \eta(1-p_d/2)\ket{{1_{\mathcal{X}}}}\bra{{1_{\mathcal{X}}}} +\eta p_d/2\ket{{0_{\mathcal{X}}}}\bra{{0_{\mathcal{X}}}} +\\+ (1-\eta)p_d(1-p_d/2)\mathds{1}, \\\hat{\Gamma}_{f} = (1-p_d)^2(1-\eta)\mathds{1},
\end{gathered}
\end{equation}
After omitting the $p_d^2$ terms, we obtain
\begin{equation}
    \begin{gathered}
     \hat{\Gamma}_{0_{\mathcal{X}}} = \eta(1-p_d/2)\ket{{0_{\mathcal{X}}}}\bra{{0_{\mathcal{X}}}} +\eta p_d/2\ket{{1_{\mathcal{X}}}}\bra{{1_{\mathcal{X}}}} +\\+ (1-\eta)p_d\mathds{1}, \\\hat{\Gamma}_{1_{\mathcal{X}}} = \eta(1-p_d/2)\ket{{1_{\mathcal{X}}}}\bra{{1_{\mathcal{X}}}} +\eta p_d/2\ket{{0_{\mathcal{X}}}}\bra{{0_{\mathcal{X}}}} +\\+ (1-\eta)p_d\mathds{1}, \\\hat{\Gamma}_{f} = (1-2p_d)(1-\eta)\mathds{1}.
\end{gathered}
\end{equation}
The last step is to extend this reasoning to cases beyond the qubit assumption. In every imperfection model considered in this work, we assume that Bob expects the incoming state to appear in a two-dimensional encoding subspace. With the above form of the POVM it is easy to then limit our measurements to just this subspace - we simply need to replace the projections $\ket{\gamma}\bra{\gamma}$ with the projections onto the encoding qubit subspace, and the identities $\mathds{1}$ become identities acting on the entire Hilbert space. We need to also add a failure term corresponding to the probability of the pulse arriving in the Hilbert space not measured by Bob's detector. Due to the difference in orders of magnitude, we decided to omit the dark count contributions $\eta p_d \mathds{1}_{side\:channel}$ that appear when the pulse arrives to the detector but is not measured after the projection, since for most distances considered it is very small compared to the standard dark count term $(1-\eta)p_d \mathds{1}$. After that modification, the POVM becomes
\begin{equation}
    \begin{gathered}
    \hat{\Gamma}_{0_{\mathcal{X}}} = \eta(1-p_d/2)\ket{{0_{\mathcal{X}}}}\bra{{0_{\mathcal{X}}}}_{encoding} +\\+\eta p_d/2\ket{{1_{\mathcal{X}}}}\bra{{1_{\mathcal{X}}}}_{encoding} +(1-\eta)p_d \mathds{1}, \\\hat{\Gamma}_{1_{\mathcal{X}}} = \eta(1-p_d/2)\ket{{1_{\mathcal{X}}}}\bra{{1_{\mathcal{X}}}}_{encoding} +\\+\eta p_d/2\ket{{0_{\mathcal{X}}}}\bra{{0_{\mathcal{X}}}}_{encoding} +(1-\eta)p_d \mathds{1}, \\\hat{\Gamma}_{f} = (1-2p_d)(1-\eta)\mathds{1} + \eta\mathds{1}_{side\:channel} .
     \label{single_pulse_povm}
\end{gathered}
\end{equation}

In order to compute the statistics for a pulse labeled $k$ within a block of $m$ states, we simply extend the POVM by computing a tensor product with the identities acting on Hilbert spaces of all other pulses. This produces the POVM
\begin{equation}
    \begin{gathered}
     \hat{\Gamma}_{0_{\mathcal{X}}}^k = \{\eta(1-p_d/2)\ket{{0_{\mathcal{X}}}}\bra{{0_{\mathcal{X}}}}_{k_{encoding}} +\\+\eta p_d/2\ket{{1_{\mathcal{X}}}}\bra{{1_{\mathcal{X}}}}_{k_{encoding}} + (1-\eta)p_d\mathds{1}_k\}\bigotimes_{l\neq k} \mathds{1}_l , \\\hat{\Gamma}_{1_{\mathcal{X}}}^k = \{\eta(1-p_d/2)\ket{{1_{\mathcal{X}}}}\bra{{1_{\mathcal{X}}}}_{k_{encoding}} +\\+\eta p_d/2\ket{{0_{\mathcal{X}}}}\bra{{0_{\mathcal{X}}}}_{k_{encoding}} + (1-\eta)p_d\mathds{1}_k + \}\bigotimes_{l\neq k}\mathds{1}_l, \\\hat{\Gamma}_{f}^k = \{(1-2p_d)(1-\eta)\mathds{1}_k + \eta\mathds{1}_{k_{side\:channel}}\}\bigotimes_{l\neq k}\mathds{1}_l,
     \label{pulse_k_povm}
\end{gathered}
\end{equation}
where the $k_{encoding}$ and $k_{side\:channel}$ subspaces are the encoding and side channel subspaces for pulse $k$, and $\mathds{1}_l$ is the identity operator acting on the Hilbert space corresponding to pulse $l$.

Now, to compute the statistic in the i.i.d case, we simply take the set $\{\ket{\psi_i}\}_i$ and for each state from that state we assign
\begin{equation}
    \bra{\psi_i}\hat{\Gamma}_{\gamma}\ket{\psi_i} = Y^{\gamma}_i.
\end{equation}
These statistics then correspond to a situation where the noise or errors in the channel are caused by natural phenomena of loss, detector efficiency and dark counts. For the block case, we simply compute
\begin{equation}
    \bra{\psi_{\vec{i}}}\hat{\Gamma}_{\gamma}^k\ket{\psi_{\vec{i}}} = Y^{\gamma,k}_{\vec{i}}.
\end{equation}
and we can use the above statistics directly to formulate the constraints for Model II or average over $\vec{i}$ and $k$ to calculate the statistics for Model III.

\subsection*{Simulating the QBER for a given imperfect QKD protocol}
Simulation of the quantum bit error rate (QBER) is very similar to the calculation of the mismatched basis statistics discussed in the previous section. We start by formulation a $\mathcal{Z}$ basis POVM analogous to the  $\mathcal{X}$ basis POVM from equation \ref{pulse_k_povm}
\begin{equation}
    \begin{gathered}
     \hat{\Gamma}_{0_{\mathcal{Z}}}^k = \{\eta(1-p_d/2)\ket{{0_{\mathcal{Z}}}}\bra{{0_{\mathcal{Z}}}}_{k_{encoding}} +\\+\eta p_d/2\ket{{1_{\mathcal{Z}}}}\bra{{1_{\mathcal{Z}}}}_{k_{encoding}} + (1-\eta)p_d\mathds{1}_k\}\bigotimes_{l\neq k} \mathds{1}_l , \\\hat{\Gamma}_{1_{\mathcal{Z}}}^k = \{\eta(1-p_d/2)\ket{{1_{\mathcal{Z}}}}\bra{{1_{\mathcal{Z}}}}_{k_{encoding}} +\\+\eta p_d/2\ket{{0_{\mathcal{Z}}}}\bra{{0_{\mathcal{Z}}}}_{k_{encoding}} + (1-\eta)p_d\mathds{1}_k\}\bigotimes_{l\neq k}\mathds{1}_l, \\\hat{\Gamma}_{f}^k = \{(1-2p_d)(1-\eta)\mathds{1}_k + \\+\eta\mathds{1}_{k_{side\:channel}}+ \eta\mathds{1}_{k_{side\:channel}}\}\bigotimes_{l\neq k}\mathds{1}_l.
     \label{pulse_k_povm_Z}
\end{gathered}
\end{equation}
To derive this form, we followed the exact same reasoning as in the previous section, but using the perfect $\mathcal{Z}$ basis measurements as the starting point. These POVMs now allow us to calculate the $\mathcal{Z}$ basis statistics for each pulse $k$
\begin{equation}
     \bra{\psi_{\vec{i}}}\hat{\Gamma}_{\gamma}^k\ket{\psi_{\vec{i}}} = Y^{\gamma,k}_{\vec{i}}.
\end{equation}
 Now, for a pulse $k$, the gain, i.e. the probability of obtaining a valid detection assuming a $Z$ basis choice by Bob and Alice is
\begin{equation}
    Y_{\mathcal{Z}}^k = \frac{1}{2^m}\sum_{\vec{i}\in\{0_{\mathcal{Z}},1_{\mathcal{Z}}\}^{\otimes m}}\bra{\psi_{\vec{i}}}\hat{\Gamma}_{0_{\mathcal{Z}}}^k+\hat{\Gamma}_{1_{\mathcal{Z}}}^k\ket{\psi_{\vec{i}}}
\end{equation}
while the probability of a bit error occurring on pulse $k$ is given by
\begin{equation}
    Y_{Z \wedge bit }^k =\frac{1}{2^m}\sum_{\vec{i}\in\{0_{\mathcal{Z}},1_{\mathcal{Z}}\}^{\otimes m}}\bra{\psi_{\vec{i}}}\hat{\Gamma}_{\tilde{i}_{k}}^k\ket{\psi_{\vec{i}}}
\end{equation}
where $\tilde{i}_k$ denotes the NOT operation on the $i_k$-th entry of the $\vec{i}$ vector, that is $0_{\mathcal{Z}}$ for $i_k = 1_{\mathcal{Z}}$ and vice versa. The bit error rate on pulse $k$ will then be given by renormalizing $Y_{Z \wedge bit }^k$ by the gain $Y_{\mathcal{Z}}^k$. We can then take the average weighted by gain over $k$ to compute the error rate of the entire block.

\end{document}